\documentclass[12pt]{article}
\usepackage{amsmath,amsfonts,amssymb,latexsym,graphics}
\usepackage[driver]{graphicx}

\begin{document}

\newcommand{\R}{{\mathbb R}}
\newcommand{\C}{{\mathbb C}}
\newcommand{\N}{{\mathbb N}}
\newcommand{\Z}{{\mathbb Z}}
\newcommand{\PP}{{\mathbb P}}
\newcommand{\SSS}{{\mathbb S}}
\newcommand{\WW}{{\mathbb W}}
\newcommand{\EE}{{\mathbb E}}
\newcommand{\Llr}{{\Longleftrightarrow}}
\newcommand{\kA}{{\cal A}}
\newcommand{\kB}{{\cal B}}
\newcommand{\kC}{{\cal C}}
\newcommand{\kD}{{\cal D}}
\newcommand{\kE}{{\cal E}}
\newcommand{\kF}{{\cal F}}
\newcommand{\kG}{{\cal G}}
\newcommand{\kH}{{\cal H}}
\newcommand{\kK}{{\cal K}}
\newcommand{\kJ}{{\cal J}}
\newcommand{\kL}{{\cal L}}
\newcommand{\kM}{{\cal M}}
\newcommand{\kN}{{\cal N}}
\newcommand{\kO}{{\cal O}}
\newcommand{\kP}{{\cal P}}
\newcommand{\kR}{{\cal R}}
\newcommand{\kS}{{\cal S}}
\newcommand{\kT}{{\cal T}}
\newcommand{\kW}{{\cal W}}
\newcommand{\kX}{{\cal X}}
\newcommand{\kY}{{\cal Y}}

\newcommand{\ga}{{\alpha}}
\newcommand{\gb}{{\beta}}
\newcommand{\gd}{{\delta}}
\newcommand{\gD}{{\Delta}}
\newcommand{\gga}{{\gamma}}
\newcommand{\gG}{{\Gamma}}
\newcommand{\gF}{{\Phi}}
\newcommand{\gf}{{\phi}}
\newcommand{\gk}{{\kappa}}
\newcommand{\gK}{{\Kappa}}
\newcommand{\gL}{{\Lambda}}
\newcommand{\lam}{{\lambda}}
\newcommand{\gl}{{\lambda}}
\newcommand{\gm}{{\mu}}
\newcommand{\gn}{{\nu}}
\newcommand{\go}{{\omega}}
\newcommand{\gO}{{\Omega}}
\newcommand{\gP}{{\Pi}}
\newcommand{\gS}{{\Sigma}}
\newcommand{\gr}{{\rho}}
\newcommand{\gs}{{\sigma}}
\newcommand{\gt}{{\theta}}
\newcommand{\gT}{{\Theta}}
\newcommand{\gU}{{\Upsilon}}
\newcommand{\Xii}{{\Sigma}}
\newcommand{\gx}{{\xi}}
\newcommand{\gX}{{\Sigma}}
\newcommand{\gY}{{\Upsilon}}
\newcommand{\gz}{{\zeta}}
\newcommand{\f}{{\varphi}}
\newcommand{\ff}{{\varphi}}
\newcommand{\e}{{\varepsilon}}
\newcommand{\ee}{{\varepsilon}}
\newcommand{\btr}{\Pi=\{{\cal H},\Gamma_0,\Gamma_1\}}
\newcommand{\gotC}{\mathfrak C}
\newcommand{\gotb}{\mathfrak b}
\newcommand{\gotH}{{\mathfrak H}}
\newcommand{\gotS}{\mathfrak S}
\newcommand{\gotK}{{\mathfrak H}}
\newcommand{\gotR}{\mathfrak R}

\newcommand{\real}{{\Re\mbox{\rm e}}}
\newcommand{\imag}{{\Im\mbox{\rm m}}}
\newcommand{\dom}{{\mbox{\rm dom}}}
\newcommand{\sign}{{\mbox{\rm sign}}}
\newcommand{\supp}{\mbox{\rm supp}}
\newcommand{\mes}{\mbox{\rm mes}}
\newcommand{\diag}{\mbox{\rm diag}}
\newcommand{\cl}{\mbox{\rm cl}}
\newcommand{\ran}{{\mbox{\rm ran}}}
\newcommand{\wt}[1]{{\widetilde{#1}}}
\newcommand{\Ext}{{\mbox{\rm Ext}}}

\newtheorem{thm}{\hspace{45pt}THEOREM}
\newtheorem{theorem}[thm]{\hspace{45pt}THEOREM}
\newtheorem{proposition}[thm]{\hspace{45pt}PROPOSITION}
\newtheorem{lemma}[thm]{\hspace{45pt}LEMMA}
\newtheorem{cor}[thm]{\hspace{45pt}COROLLARY}
\newtheorem{ex}[thm]{\hspace{45pt}EXAMPLE}
\newtheorem{definition}[thm]{\hspace{45pt}DEFINITION}
\newtheorem{remark}[thm]{\hspace{45pt}REMARK}

\renewcommand{\thesection}{\Roman{section}}

\newenvironment{proof}%
{\begin{sloppypar}{\bf PROOF.}}%
{\hspace*{\fill}$\square$\end{sloppypar}}

\newenvironment{proof1}
{\begin{sloppypar}{\bf PROOF OF THEOREM 1.1.}}%
{\hspace*{\fill}$\square$\end{sloppypar}}

\renewcommand{\thefootnote}{\fnsymbol{footnote}}

\title{\bf 
CONVERGENCE OF SCHR{\"O}DINGER OPERATORS}

\setcounter{footnote}{0}
\renewcommand{\thefootnote}{\arabic{footnote}}

\vspace{1cm}

\author{\normalsize
JOHANNES F. BRASCHE$^{a,b}$ and KATE\v{R}INA O\v{Z}ANOV\'{A}$^b$}
\date{\today}

\renewcommand{\thefootnote}{\fnsymbol{footnote}}

\setcounter{footnote}{0}
\renewcommand{\thefootnote}{\arabic{footnote}}

\maketitle 

\begin{quote}
{\small \em a) Institute of Mathematics, TU Clausthal,
38678 Clausthal-Z., Germany \\
b) Department of Mathematics, CTH \mbox{\footnotesize{$\&$}} GU,
41296 G{\"o}teborg, Sweden \\ \phantom{a)} 
\rm brasche@math.chalmers.se, nemco@math.chalmers.se}
\end{quote}
\vspace{1cm}
\begin{quote}
{\bf Abstract}: 
We prove two limit relations between Schr\"{o}dinger operators perturbed
by measures. First, weak convergence of finite real-valued Radon measures 
$\mu_n \longrightarrow m$ implies that the operators 
$-\gD + \ee^2 \gD^2 + \mu_n$ in $L^2(\R^d,dx)$ converge to 
$-\gD + \ee^2 \gD^2 + m$ in the norm resolvent sense, provided $d\le 3$.
Second, for a large family, including the Kato class,
of real-valued Radon measures $m$, the operators
$-\gD + \ee^2 \gD^2 + m$ tend to the operator
$-\gD +m$ in the norm resolvent sense as $\ee$ tends to zero. 
Explicit upper bounds for the rates of convergences are derived. 
Since one can choose point measures $\mu_n$ with mass at only finitely many
points, a combination of both convergence results leads to 
an efficient method for the numerical computation of the
eigenvalues in the discrete spectrum and corresponding eigenfunctions
of Schr{\"o}dinger operators. The approximation is illustrated by numerical
calculations of eigenvalues for one simple example of measure $m$.

\end{quote}

\section{Introduction}

In this paper we are going to analyze convergence of Schr{\"o}dinger 
operators perturbed by measures. It is known that
weak convergence of potentials implies norm-resolvent convergence
of the corresponding one-dimensional Schr{\"o}dinger operators. This 
result from \cite{bft} may be interesting for several reasons. For
instance every finite real-valued Radon measure on $\R$ 
is the weak limit of a sequence of point measures with mass
at only finitely many points. There exist efficient numerical methods for the
computation of the eigenvalues and corresponding eigenfunctions
of one-dimensional Schr{\"o}dinger operators with a potential supported
by a finite set; actually the effort for the computation grows at most 
linearly with the number of points of the support \cite{m}. Since the
norm resolvent convergence implies convergence of the eigenvalues in the  
discrete spectra  and corresponding eigenspaces, we get an efficient method 
for the numerical calculation of the discrete spectra of one-dimensional 
Schr{\"o}dinger operators. Norm resolvent convergence has also other 
important consequences: locally uniform convergence of the associated unitary 
groups and semigroups, convergence of the spectral projectors (which implies 
the mentioned results on the discrete spectra) etc.

Let us also mention a completely different motivation for studying convergence
of operators with point potentials. In quantum mechanics
neutron scattering is often described via so called zero-range 
Hamiltonians (the monograph \cite{aghh} is an excellent standard reference
to this research area). In a wide variety of models the positions of the
neutrons are described via a family $(X_j)_{j=1}^n$ of independent
random variables with joint distribution $\mu$. 
Usually the number $n$ of neutrons is large and one is interested in the
limit when $n$ tends to infinity and the strengths of the single size 
potentials tend to zero. In the one-dimensional case
this motivates to investigate the limits of operators of the form
$$ 
-\frac{d^2}{dx^2} + \frac{a}{n} \sum_{j=1}^n \delta_{X_j(\go)},
\qquad \go \in \gO,
$$
$a\not= 0$ being a real constant and $(\gO,\kF,\PP)$ a probability space. 
By the theorem of Glivenko-Cantelli, 
for $\PP$-almost all $\go\in \gO$ the sequence 
$ ( \frac{a}{n} \sum_{j=1}^n \delta_{X_j(\go)})_{n\in \N}$ converges
to the measure $a\mu$ weakly. By the mentioned
result from \cite{bft}, this implies that
$$ 
-\frac{d^2}{dx^2} + a\mu = \lim_{n\longrightarrow \infty}  
\left( -\frac{d^2}{dx^2} + \frac{a}{n} 
\sum_{j=1}^n \delta_{X_j(\go)} \right)
$$
in the norm resolvent sense $\PP$-a.s. 

It is the purpose of the present note to derive analogous results in the
two- and three-dimensional case. It was shown in \cite{bft} and \cite{en}
that one can approximate Schr{\"o}dinger operators perturbed by suitable
measures by point potential Hamiltonian. However, the convergence there
was in the strong resolvent sense, which is of course a weaker result than
the norm resolvent convergence. 

If the dimension is higher than one, then it seems to be impossible to work 
directly with operators of the form $-\gD + \mu$, $\mu$ being a point measure. 
In fact, while the operators 
$-\frac{d^2}{dx^2} + \sum_{j=1}^n a_j \delta_{x_j}$ 
can be defined in dimension one via Kato's quadratic form method as the 
unique lower semibounded self-adjoint operator associated to the energy form 
\begin{eqnarray*}
D(\kE) & := & H^1(\R),\\
\kE (f,f) & := & \int \vert f'(x) \vert^2 dx + 
\sum_{j=1}^n a_j \vert \tilde{f}(x_j) \vert^2,\quad f\in D(\kE),
\end{eqnarray*}
$\tilde{f}$ being the unique continuous representative of $f\in H^1(\R)$, 
in higher dimension $d > 1$, the quadratic form 
\begin{eqnarray*}
D(\kE) & := & \{ f\in H^1(\R^d): f\mbox{ has a continuous representative } 
\tilde{f} \},\\
\kE (f,f) & := & \int \vert \nabla f(x) \vert^2 dx + 
\sum_{j=1}^n a_j \vert \tilde{f}(x_j) \vert^2,\quad f\in D(\kE),
\end{eqnarray*}
is not lower semibounded and closable if at least one coefficient
$a_j$ is different from zero.

The strategy to overcome the mentioned problem in higher dimensions is based
on two simple observations:

\noindent
1. The lower semibounded self-adjoint operator 
$\gD^2 + \mu$ can be defined via Kato's quadratic form method 
for every real-valued finite Radon measure
$\mu$ on $\R^d$ (if $d\in \{ 1,2,3 \}$), including point measures.

\noindent
2. $-\gD + \ee^2 \gD^2 \longrightarrow -\gD$ in the norm resolvent sense, as
$\ee>0$ tends to zero.

We show the convergence claim in two steps. In section \ref{sec:mu} we 
shall prove that the sequence $(-\gD + \ee^2 \gD^2 + \mu_n)_{n\in \N}$ 
converges to $-\gD + \ee^2 \gD^2 + m$  in the norm resolvent sense provided 
$d\le 3$, $\ee >0$ and the finite real-valued Radon measures $\mu_n$ on $\R^d$ 
converge to the finite real-valued Radon measure $m$ weakly. Then, 
for a large class of measures $m$ we shall prove that 
$$ -\gD + \ee^2 \gD^2 + m \longrightarrow -\gD + m$$
in the norm resolvent sense as $\ee$ tends to zero, cf. section
\ref{sec:lime}. 
Actually, we will not only prove convergence but also give explicit
error estimates.

As approximating measures $\mu_n$ we can, in particular, choose
point measures with mass at only finitely many points. 
In section \ref{sec:eig} we will present formulae which make it possible 
to calculate the eigenvalues and corresponding eigenspaces 
of operators perturbed by a finite number point measures. Then similarly
to \cite[chapter II.2]{aghh}, the spectral problem means to solve an
implicit equation and the effort for these computations grows at most as 
$\kO(n^3)$. 

Putting both convergence results from sections \ref{sec:mu} and \ref{sec:lime}
and formulae from section \ref{sec:eig} together,  
we get an efficient method to calculate the eigenvalues in the discrete 
spectrum and corresponding eigenspaces of Schr{\"o}dinger operators
$-\gD + m$ numerically. We apply the approximation to the simple two-dimensional
example, where measure $m$ is negative and supported by a circle.

Our method does not only cover the case when $m$ is absolutely 
continuous w.r.t. the $(d-1)$-dimensional volume measure of a manifold
with codimension one but a fairly large class of 
measures $m$ containing the set of all finite real-valued measures belonging
to the Kato class. In particular, the absolutely continuous case 
$dm = Vdx$ where $-\gD+m = -\gD+V$ is a regular
Schr{\"o}dinger operator is contained in our approach. We refer to 
\cite{pos} for related convergence results in the regular case.

{\bf Notation and auxiliary results}: 
\noindent
Let $\mu$ be a real-valued Radon measure on $\R^d$. By the 
Hahn-Jordan theorem, there exist unique positive Radon measures 
$\mu^{\pm}$ on $\R^d$ such that 
\begin{eqnarray}
\mu= \mu^+ - \mu^- \mbox{ and } \mu^+(\R^d\setminus B) = 0 = \mu^-(B)
\nonumber
\end{eqnarray}
for some suitably chosen Borel set $B$. We put
\begin{eqnarray}
\parallel \mu \parallel := \mu^+ (\R^d) + \mu^-(\R^d)\mbox{ and }
\vert  \mu \vert := \mu^+ + \mu^-. \nonumber
\end{eqnarray}
If $\mu$ is finite, then we define its Fourier transform $\hat{\mu}$ as
$$ 
\hat{\mu}(p) := (2\pi)^{-d/2} \int e^{ipx} \mu(dx),\qquad p\in \R^d.
$$
Similarly, $\hat{f}$ also denotes the Fourier transform of 
$f\in L^2(dx):=L^2(\R^d,dx)$, $dx$ being the Lebesgue measure.

For $s>0$ we denote the Sobolev space of order $s$ by $H^s(\R^d)$, i.e.
\begin{eqnarray*}
H^s(\R^d) & := & \left\{ f\in L^2(dx): \, 
\int (1+ p^2)^s \vert \hat{f} (p) \vert^2 dp < \infty \right\}, \\
\parallel f \parallel_{H^s} & := & 
\left(\int (1+ p^2)^s \vert \hat{f} (p) \vert^2 dp \right)^{1/2},
\qquad f\in H^s(\R^d).
\end{eqnarray*}
We shall use occasionally the abbreviations $L^2(\mu) := L^2(\R^d,\mu)$ 
and $H^s:= H^s(\R^d)$. 

$\parallel T \parallel_{\kH_1,\kH_2}$ denotes the operator norm of 
$T$ as an operator from $\kH_1$ to $\kH_2$ and \newline
$\parallel T \parallel_{\kH} := \parallel T \parallel_{\kH,\kH}$.
$\parallel f \parallel_{\kH}$ and $(f,h)_{\kH}$ represent 
the norm and the scalar product in the Hilbert $\kH$, respectively. 
If the reference to a measure is missing, then we tacitly refer 
to the Lebesgue measure $dx$. 
For instance ``integrable'' means ``integrable w.r.t. $dx$'' if not 
stated otherwise, $\parallel T \parallel$, $(f, h)$ and 
$\parallel f \parallel$ denote the operator norm of $T$,
scalar product and norm in the Hilbert space $L^2(dx)$, respectively. 
We denote by $C_0^{\infty}(\R^d)$ the space of smooth functions with 
compact support.

For arbitrary $\ee\ge 0$ ($\ee =0$ will be admitted only in section 
\ref{sec:lime}) let  $\kE_{\ee}$ be the nonnegative closed
quadratic form in the Hilbert space $L^2(dx)$ associated to the 
nonnegative self-adjoint operator $-\gD + \ee^2 \gD^2 $ in $L^2(dx)$.
Obviously we have 
\begin{eqnarray}
D(\kE_{\ee}) & = & H^2(\R^d), \nonumber \\
\kE_{\ee} (f,f) & = & \ee^2 \, (\gD f, \gD f) + (\nabla f,\nabla f)
\ge \ee^2\, (\gD f, \gD f),\qquad f\in D(\kE_{\ee}), \nonumber
\end{eqnarray}
for every $\ee >0$. Note that for $\ee=0$ the form domain is $H^1(\R^d)$
and $\kE_0$ is the classical Dirichlet form.
For any $\alpha >0$ we put 
\begin{eqnarray}
\kE_{\ee,\alpha} (f,h) := \kE_{\ee} (f,h) + \alpha (f,h),\qquad 
f,h\in D(\kE_{\ee}).\nonumber
\end{eqnarray}
%

\section{Operator norm convergence}\label{sec:mu}

Throughout this section let $d\le 3$ and $\mu$ be a finite real-valued Radon
measure on $\R^d$. Then, by Sobolev's 
embedding theorem, for every $s>3/2$, and, in particular, for $s=2$, 
every $f\in H^s(\R^d)$ has a unique continuous
representative $\tilde{f}$ and 
\begin{eqnarray}\label{1.2}
\parallel \tilde{f} \parallel_{\infty} := 
\sup \{ \vert \tilde{f} (x) \vert : \, x\in \R^d\} \le c_s  
\parallel f \parallel_{H^s},\qquad f\in H^s(\R^d), 
\end{eqnarray}
for some finite constant $c_s$. Note that $c_s\le 1$ if $s=2$. 
It follows that for every $\ee>0$ and every $\eta >0$ there
exists an $\alpha = \alpha(\ee,\eta) <\infty$ such that
\begin{eqnarray}
\parallel \tilde{f} \parallel_{\infty}^2 \le \eta\, \kE_{\ee}(f,f)
+ \alpha (f,f),\qquad f\in H^2(\R^d).
\end{eqnarray}
Since $\mu$ is finite, for arbitrary $\ee,\eta>0$ and some finite $\alpha$
we get 
\begin{eqnarray}\label{1.6}
\vert \int \vert \tilde{f} \vert^2 d\mu \vert \le 
\eta \parallel \mu \parallel \kE_{\ee} (f,f) + 
\alpha \parallel \mu \parallel (f,f),\qquad f\in H^2(\R^d). 
\end{eqnarray}

We put 
\begin{eqnarray*}
D(\kE_{\ee}^{\mu}) & := & H^2(\R^d), \\
\kE_{\ee}^{\mu} (f,f) & := & \kE_{\ee} (f,f) + 
\int \vert \tilde{f} \vert^2 d\mu,\qquad f\in  D(\kE_{\ee}^{\mu}). 
\end{eqnarray*}
By (\ref{1.6}) and the KLMN-theorem, $\kE_{\ee}^{\mu}$ is a lower
semibounded closed quadratic form in $L^2(dx)$. We denote the
lower semibounded self-adjoint operator in $L^2(dx)$ associated to
$\kE_{\ee}^{\mu}$ by $-\gD + \ee^2 \gD^2 + \mu$.

Our main tool to prove convergence results will be a Krein-like formula
which expresses the resolvent $(-\gD + \ee^2 \gD^2 + \mu + \alpha)^{-1}$ 
by means of the resolvent
$$
G_{\ee,\alpha} := (-\gD + \ee^2 \gD^2 + \alpha)^{-1}.
$$
The operator $G_{\ee,\alpha}$ has the integral 
kernel $g_{\ee,\alpha}(x-y)$  with Fourier transform 
\begin{eqnarray*}
\hat{g}_{\ee,\alpha}(p) := \frac{1}{\ee^2 p^4+ p^2 +\alpha}, \qquad
p \in \R^d.
\end{eqnarray*}
For every $\ee \ge 0$ and $\alpha >0$, the
function $g_{\ee,\alpha}(x)$ is continuous on $\R^d \setminus \{ 0\}$ and
if $d=1$ or if $d\le 3$ and $\ee >0$ it is continuous on whole $\R^d$.
Moreover, it is radially symmetric. Finally, $g_{0,\alpha}$ is the Green 
function of the free Laplacian in $\R^d$ and it is nonnegative. 
By the dominated convergence theorem, 
\begin{eqnarray}\label{G1}
\parallel g_{\ee,\alpha} \parallel_{H^2}^2 = 
\int \frac{(1+ p^2)^2}{ \vert \ee^2 p^4 + p^2 +\alpha \vert^2} \,dp
\longrightarrow 0, \qquad \mbox{as } \vert \alpha \vert \longrightarrow \infty
\end{eqnarray}
which, by Sobolev's inequality, implies that
\begin{eqnarray}\label{G2}
\parallel g_{\ee,\alpha} \parallel_{\infty} 
\longrightarrow 0, \qquad \mbox{as } \vert \alpha \vert \longrightarrow \infty.
\end{eqnarray}

The fact that $g_{\ee,\alpha}$ is the Green function of $-\gD +\ee^2 \gD^2$
means that
\begin{eqnarray}
\int  g_{\ee,\alpha}(x-y)
(-\gD + \ee^2 \gD^2 +\alpha) h(y) dy = h(x)
\qquad dx\mbox{-a.e.} \nonumber
\end{eqnarray}
for all $h\in D(-\gD + \ee^2 \gD^2)=H^4(\R^d)$. The equation above does not 
only hold almost everywhere w.r.t. the Lebesgue measure $dx$ but even 
pointwise everywhere, as the following lemma states.
\begin{lemma}\label{lem: cont}
Let Green function $g_{\ee,\alpha}$ and operator $-\gD + \ee^2 \gD^2 + 
\alpha$ be defined as above. Then one has
\begin{eqnarray}\label{1.10a}
\int  g_{\ee,\alpha}(x-y) (-\gD + \ee^2 \gD^2 + \alpha) h(y) dy 
= \tilde{h}(x),
\qquad x\in \R^d
\end{eqnarray}
for all $h\in H^4(\R^d)$.
\end{lemma}
{\bf Proof}: In fact, we have only to show that the integral on the left hand 
side is a continuous function of $x \in \R^d$. We choose any sequence 
$(f_n)_{n\in \N}$ of continuous functions with compact support converging to 
$ (-\gD + \ee^2 \gD^2 +\alpha) h$ in $L^2(dx)$. By (\ref{G1}), 
$g_{\ee,\alpha}\in H^2(\R^d) \subset  L^2(dx)$, therefore we can write
\begin{eqnarray}
\int  g_{\ee,\alpha}(x-y) (-\gD + \ee^2 \gD^2 +\alpha) h(y) dy 
= \lim_{n\longrightarrow \infty} \int  g_{\ee,\alpha}(x-y)
f_n(y) dy , \qquad x\in \R^d. \nonumber
\end{eqnarray}
Obviously the mapping $x\mapsto \int  g_{\ee,\alpha}(x-y)
f_n(y) dy$, $\R^d \longrightarrow \C$, is 
the unique continuous representative $\widetilde{G_{\ee,\alpha}f_n}$ 
of $G_{\ee,\alpha} f_n$
for every
$n\in \N$. Since $G_{\ee,\alpha}$ is a bounded operator from
$L^2(dx)$ to $H^2(\R^d)$ (even to $H^4(\R^d)$),  the sequence 
$(G_{\ee,\alpha} f_n)_{n\in \N}$ converges in $H^2(\R^d)$ to
$G_{\ee,\alpha}(-\gD + \ee^2 \gD^2 +\alpha) h =h$. 
By Sobolev's inequality (\ref{1.2}), this implies that the sequence 
$(\widetilde{G_{\ee,\alpha}f_n})_{n\in \N}$ 
of the unique continuous representatives 
converges to a continuous function uniformly. By the last equality,
$x\mapsto \int g_{\ee,\alpha} (x-y) (-\gD + \ee^2 \gD^2 + \alpha) h(y) dy$,
 $\R^d \longrightarrow \C$, is this continuous uniform limit and we have
proved (\ref{1.10a}). 
\hfill $\Box$

We introduce following integral operator
\begin{eqnarray}
G_{\ee,\alpha}^{\mu} f(x) := \int g_{\ee,\alpha} (x-y) \tilde{f}(y) \mu(dy)
\qquad dx\mbox{-a.e., }f\in H^2(\R^d).\nonumber
\end{eqnarray}
We can prove several estimates of its operator
norm.
\begin{lemma}\label{lem: ineq}
The operator $G_{\ee,\alpha}^{\mu}$ is bounded on $H^2(\R^d)$ and its operator
norm $\parallel G^{\mu}_{\ee,\alpha} \parallel_{H^2}$ 
decays with $\alpha \longrightarrow \infty$. The operator is bounded
also w.r.t. other operator norms, in particular there are finite real
numbers $c_i$, $i=1,2,3$ such that
\begin{eqnarray*}
\parallel G_{\ee,\alpha}^{\mu} f \parallel_{H^2} &\le& c_1(\alpha) \parallel 
\tilde{f} \parallel_{\infty} \\
\parallel G_{\ee,\alpha}^{\mu} f \parallel_{L^2} &\le& c_2(\alpha) \parallel 
\tilde{f} \parallel_{L^2(|\mu|)}  \qquad f \in H^2(\R^d)\\
\parallel \widetilde{G_{\ee,\alpha}^{\mu} f} \parallel_{L^2(|\mu|)} &\le& 
c_3(\alpha) \parallel \tilde{f} \parallel_{L^2(|\mu|)}
\end{eqnarray*}
and all three numbers $c_i$ vanish in the limit 
$\alpha \longrightarrow \infty$.
\end{lemma}
{\bf Proof}: Using Sobolev's inequality we have for arbitrary $f\in H^2(\R^d)$
\begin{eqnarray}
\vert \widehat{\tilde{f}\mu} (p) \vert^2 \le (2\pi)^{-d}
\parallel \tilde{f} \parallel_{\infty}^2 \parallel \mu \parallel^2 \le
(2\pi)^{-d} \parallel \tilde{f} \parallel_{H^2}^2 \parallel \mu \parallel^2,
\qquad p\in \R^d. \nonumber
\end{eqnarray}
Then the convolution theorem yields
\begin{eqnarray}
\parallel   G_{\ee,\alpha}^{\mu} f         \parallel_{H^2}^2
& = & \int \vert (1+p^2)^2 \vert \,
\vert ( {g_{\ee,\alpha} * \tilde{f}\mu} \hat{)}
(p) \vert^2 dp\nonumber \\
& = & (2\pi)^d \int \frac{(1+p^2)^2}{\vert \ee^2 p^4 + p^2 + \alpha \vert^2} \,
\vert \widehat{\tilde{f}\mu}(p)\vert^2 dp \nonumber \\
& \le & \int \frac{(1+p^2)^2}{\vert \ee^2 p^4 + p^2 + \alpha \vert^2} \,
\parallel \tilde{f} \parallel_{\infty}^2  
\parallel \mu \parallel^2  dp \nonumber \\
& \le & \int \frac{(1+p^2)^2}{\vert \ee^2 p^4 + p^2 + \alpha \vert^2}\, dp\,
\parallel \tilde{f} \parallel_{H^2}^2  \parallel \mu \parallel^2 < 
\infty,\qquad f\in H^2(\R^d). \nonumber
\end{eqnarray}
Therefore $G_{\ee,\alpha}^{\mu}$ is an everywhere defined bounded 
operator on $H^2(\R^d)$ and we get an upper bound for the norm 
\begin{eqnarray}\label{1.15}
\parallel G_{\ee,\alpha}^{\mu} \parallel_{H^2,H^2}
\le \parallel \mu \parallel \left( 
\int \frac{(1+p^2)^2}{\vert \ee^2 p^4 + p^2 + \alpha \vert^2} dp \right)^{1/2},
\end{eqnarray}
and the expression on the r.h.s. is also the uniform upper bound $c_1$.

To determine the remaining upper bounds $c_2$ and $c_3$, we can write 
\begin{eqnarray}\label{G3}
& &\int \vert G_{\ee,\alpha}^{\mu} f (x) \vert^2 dx \nonumber \\
= & & \int \vert \int g_{\ee,\alpha} (x-y) \tilde{f} (y) \mu^+(dy) 
-\int g_{\ee,\alpha} (x-y) \tilde{f}(y) \mu^- (dy) \vert^2 dx \nonumber \\
\le & & 
2 \int \vert \int g_{\ee,\alpha} (x-y) \tilde{f}(y) \mu^+ (dy)\vert^2 dx
+ 2 \int \vert \int g_{\ee,\alpha} (x-y) \tilde{f}(y) \mu^- (dy) \vert^2 dx
\nonumber \\
\le & & 
2 \int \int \vert g_{\ee,\alpha} (x-y) \vert^2 \mu^+(dy)\,
\int \vert \tilde{f} (y) \vert^2 \mu^+ (dy)\, dx\nonumber \\  
& + & 
2 \int \int \vert g_{\ee,\alpha} (x-y) \vert^2 \mu^-(dy)\,
\int \vert \tilde{f} (y) \vert^2 \mu^- (dy) \, dx\nonumber \\  
\le & & 2 \int \vert g_{\ee,\alpha}(x) \vert^2 dx \, \parallel \mu \parallel
\, \int \vert \tilde{f} (y) \vert^2 \vert \mu \vert (dy),
\qquad f\in H^2(\R^d).
\end{eqnarray}
In a similar way we arrive at
\begin{eqnarray*}
\int \vert \widetilde{G_{\ee,\alpha}^{\mu} f} (x) \vert^2 \vert \mu \vert(dx)
\le 2 \parallel g_{\ee,\alpha}  \parallel_{\infty}^2 
\parallel  \mu    \parallel^2
\int \vert \tilde{f} (y) \vert^2 \vert \mu \vert (dy).
 \end{eqnarray*}
Finally, from (\ref{G1}) and (\ref{G2}) one concludes that all the upper 
bounds of the operator norms tend to zero in the limit $\alpha \longrightarrow
\infty$. \hfill $\Box$ 

General results of \cite{bpeking} (cf. also section \ref{sec:lime} below)
provide, in particular, an explicit formula for the resolvent of the operator 
$-\gD + \ee^2 \gD^2 + \mu$. In this resolvent formula there occur operators
acting in different Hilbert spaces. This is inconvenient when we
investigate the convergence of sequences of such operators and 
we shall use a slightly different resolvent formula:
\begin{eqnarray}\label{1.17}
 (-\gD + \ee^2 \gD^2 + \mu + \alpha)^{-1} = 
 G_{\ee,\alpha}-  G_{\ee,\alpha}^{\mu} 
(I+  G_{\ee,\alpha}^{\mu})^{-1}  G_{\ee,\alpha}.
\end{eqnarray}
For the sake of completeness we present the proof of the above Krein's formula
in the appendix.
According to lemma \ref{lem: ineq}, we can choose $\alpha >0$ such
that $\parallel G_{\ee,\alpha}^{\mu} \parallel_{H^2,H^2} <1$. 
Then the operator $I+ G_{\ee,\alpha}^{\mu}$ is invertible and its
inverse is everywhere defined on $H^2(\R^d)$ and bounded; here $I$ denotes
the identity on $H^2(\R^d)$. By (\ref{1.6}), we can choose $\alpha >0$ such
that, in addition, 
\begin{eqnarray}\label{1.17a}
\kE_{\ee,\alpha}^{\mu}(f,f):= 
\kE_{\ee}^{\mu}(f,f) + \alpha (f,f) \ge (f,f), \quad f\in D(\kE_{\ee}^{\mu}).
\end{eqnarray}

We are now prepared for the proof of the main theorem of this section:
\begin{theorem}\label{thm1}
Let $m$ and $\mu_n$, $n\in \N$, be finite real-valued Radon measures on
$\R^d$. Suppose that the sequence $(\mu_n)_{n\in \N}$ converges
to $m$ weakly and $\sup_{n\in \N}  \parallel \mu_n \parallel < \infty$.
Let $\ee,\alpha >0$ and $d\in \{ 1,2,3 \}$. Then
the operators $-\Delta + \ee^2 \Delta^2 + \mu_n$ converge 
to $-\Delta + \ee^2 \Delta^2 + m$ in the 
norm resolvent sense. 
\end{theorem}
{\bf Proof}: Let $\ee>0$ be arbitrary.
We choose $0 < c < 1$ and $\alpha>0$ such that 
\begin{eqnarray}\label{1.18a}
\parallel \mu_n \parallel^2 
\int \frac{(1+p^2)^2}{\vert \ee^2 p^4 + p^2 + \alpha \vert^2} \,dp \, \le c^2,
\quad n\in \N, 
\end{eqnarray}
and 
\begin{eqnarray}\label{1.18b}
\parallel m \parallel^2 
\int \frac{(1+p^2)^2}{\vert \ee^2 p^4 + p^2 + \alpha \vert^2} \, dp \, \le c^2.
\end{eqnarray}
According to (\ref{1.6}), we can choose $\alpha >0$ such that, in addition, 
\begin{eqnarray}\label{1.18c}
\kE_{\ee,\alpha}^{\mu_n} (f,f) \ge (f,f),\qquad f\in H^2(\R^d),\quad
n\in \N.
\end{eqnarray}
Since $(\mu_n)_{n\in \N}$ converges to $m$ weakly, (\ref{1.18c}) also holds
when we replace $\mu_n$ by $m$.
By Lemma \ref{lem: ineq}, in particular estimate (\ref{1.15}), inequalities 
(\ref{1.18a}) and (\ref{1.18b}) yield
\begin{eqnarray}\label{1.18d}
\parallel G_{\ee,\alpha}^{\mu_n} \parallel_{H^2,H^2} &\le& c, \qquad n\in \N, 
\nonumber\\
\parallel G_{\ee,\alpha}^m       \parallel_{H^2,H^2} &\le& c, \\
\parallel G_{\ee,\alpha}^m f     \parallel_{H^2}     &\le& c 
\parallel \tilde{f}  \parallel_{\infty}, \qquad f\in H^2(\R^d). \nonumber
\end{eqnarray}
Hence the resolvent formula (\ref{1.17})
is valid both for $\mu = m$ and for $\mu= \mu_n$, $n\in \N$. 
By Lemma \ref{lem: ineq}, we can choose $\alpha$ sufficiently large so that 
also 
\begin{eqnarray}\label{G5}
\int \vert G_{\ee,\alpha}^m h (x) \vert^2 dx \le c^2
\int \vert \tilde{h} \vert^2 d \vert m \vert
\mbox{ and }  
\int \vert \widetilde{G_{\ee,\alpha}^m h} (x)\vert^2 \vert m \vert (dx) \le c^2
\int \vert \tilde{h} \vert^2 d \vert m \vert
\end{eqnarray}
for every $h\in H^2(\R^d)$.

For notational brevity we put
\begin{eqnarray}
g_0 := g_{0,1},\quad 
g:=  g_{\ee,\alpha},\quad 
G : = G_{\ee,\alpha},\quad G^{\mu_n} := G_{\ee,\alpha}^{\mu_n} 
\quad \mbox{and} \quad G^m := G_{\ee,\alpha}^m. \nonumber 
\end{eqnarray}
With this notation we have 
\begin{eqnarray}
& ( -\gD + \ee^2 \gD^2 + \mu_n +\alpha)^{-1} 
- ( -\gD + \ee^2 \gD^2 + m +\alpha)^{-1} 
\nonumber\\
= & G^m [ I + G^m ]^{-1}  G  - G^{\mu_n} [ I + G^{\mu_n} ]^{-1}  
G \nonumber\\
= & (G^m - G^{\mu_n} )  [ I + G^m ]^{-1}  G + 
    (G^{\mu_n} - G^m)   [ I + G^m ]^{-1} ( G^{\mu_n} - G^m ) 
[ I + G^{\mu_n} ]^{-1} G  \nonumber \\
& +  G^m   [ I + G^m ]^{-1} ( G^{\mu_n} - G^m )
[ I + G^{\mu_n} ]^{-1} G . \nonumber
\end{eqnarray}
Since $G$ is a bounded operator from $L^2(dx)$ to $H^2(\R^d)$
we have only to show that
\begin{eqnarray}\label{1.21}
\parallel G^m - G^{\mu_n}\parallel_{H^2,L^2(dx)} 
\longrightarrow 0 \qquad \mbox{as } n \longrightarrow \infty,
\end{eqnarray}
\begin{eqnarray}\label{1.22}
\parallel G^m [ I + G^m ]^{-1} ( G^m - G^{\mu_n})  
\parallel_{H^2,L^2(dx)}
\longrightarrow 0, \qquad \mbox{as } n \longrightarrow \infty.
\end{eqnarray}

We introduce 
\begin{eqnarray}
\nu_n &:=& m - \mu_n \nonumber \\ 
\nu_{nx}(dy) &:=& g(x-y) \, \nu_n(dy),\qquad x\in \R^d,\quad n\in\N. \nonumber
\end{eqnarray}
As $d\le 3$, the function
\begin{eqnarray*}
y\mapsto \int g_0 (y-a) \, f(a) \, da
\end{eqnarray*}
is continuous and bounded for every $f\in L^2(dx)$; this well known fact can 
be proved in the same way as (\ref{1.10a}). 
Since the function $g$ is bounded and $g_0$ is nonnegative it follows that
\begin{eqnarray}
\left \vert
\int \vert g(x-y) \vert \int \vert g_0(y-a) \vert \,\vert (-\gD +1) h(a)\vert 
\, da \, \nu_n^{\pm}(dy) 
\right \vert < \infty \nonumber
\end{eqnarray}
for all $x\in \R^d$ and $h\in H^2(\R^d)$. Hence by Fubini's theorem, the 
function $k_{\nu_{nx}}:\R^d\longrightarrow \R$, defined by
\begin{eqnarray}
k_{\nu_{nx}} (a) := 
\left\{ \begin{array}{ll}
\int g_0 (y-a) \, g(x-y) \, \nu_n(dy), & \quad \mbox{if defined,}\\
0, & \quad \mbox{otherwise,}
\end{array} \right. \nonumber 
\end{eqnarray}
is Borel measurable, the integral on the right hand side 
is defined and finite for almost all $a\in \R^d$ (almost
all w.r.t. the Lebesgue measure) and
\begin{eqnarray}\label{1.31}
\vert    (G^{\nu_n} h \tilde{)}  (x) \vert^2 & =& 
\vert  \int g(x-y) \, h(y) \, \nu_n(dy) \vert^2 \nonumber  \\
& = &  
\vert \int g(x-y) \int g_0 (y-a) (-\gD +1) h(a) \, da \, \nu_n(dy) 
\vert^2 \nonumber\\
& \le &  \int \vert k_{\nu_{nx}} (a) \vert^2 da \cdot 
\int \vert (-\gD +1)h (a) \vert^2 da \nonumber \\
& \le & 2 \parallel h \parallel_{H^2}^2  \int \vert k_{\nu_{nx}} (a) 
\vert^2 da,\qquad h\in H^2(\R^d), \, n\in \N.
\end{eqnarray}

Thus in order to prove (\ref{1.21}) we have only to show that
\begin{eqnarray}\label{1.32a}
\int \int \vert k_{\nu_{nx}} (a) \vert^2 da \, dx \longrightarrow 0
\qquad \mbox{as } n\longrightarrow \infty.
\end{eqnarray}
We have
\begin{eqnarray}\label{1.33}
& \int \int \vert k_{\nu_{nx}} (a) \vert^2 da \, dx =
(2\pi)^d \int \int \vert \widehat{g_0} (p) \vert^2 \vert \widehat{\nu_{nx}}(p)
\vert^2 dp \, dx \nonumber \\
= & \int \int \frac{1}{\vert 1 + p^2 \vert^2} 
\int e^{ipy} g(x-y) \, \nu_n(dy) \int e^{-ipz} g(x-z) \nu_n(dz) \, dp \, dx. 
\end{eqnarray}
Since $\vert 1 + p^2 \vert^{-2} $ and $g$ are integrable
w.r.t. the Lebesgue measure, $g$ is bounded and the Radon measures
$\nu_n$ are finite, we can change the order of integration. Let us rewrite
(\ref{1.33}) as
\begin{eqnarray*}
\int f(y,z) \, h(y,z) \, \nu_n \otimes \nu_n(dy \, dz).
\end{eqnarray*}
The function
\begin{eqnarray*}
f(y,z) := \int e^{ipy}e^{-ipz} \frac{1}{\vert 1 + p^2 \vert^2} \, dp,
\qquad y,z\in \R^d, 
\end{eqnarray*}
is bounded and continuous. It follows from the fact that it is (up to
multiplication by $(2 \pi)^{d/2}$) the inverse Fourier transform of the 
integrable function $\vert 1 + p^2 \vert^{-2}$ at the point $z-y$.

Also the function
$$  
h(y,z) := \int g(x-y) \, g(x-z) \, dx 
$$
is bounded and continuous for $y,z\in \R^d$. This can be shown using following
observation. Let $y\in \R^d$ and $K$ be any compact neighborhood of $y$. 
Since $\vert x \vert^j g_{\ee,\alpha}(x) \longrightarrow 0$ for every
$j \in N$ as $\vert x \vert \longrightarrow \infty$, there exists a 
constant $a<\infty$ such that  
$$ \vert g(x-y) \, g(x-z) \vert \le a \parallel g \parallel_{\infty}
\mbox{dist}(x,K)^{-4},\qquad x\in \R^d\setminus K,\quad 
z\in \R^d,\quad y\in K.$$

By Stone-Weierstrass theorem, the set of functions of the form 
$\sum_{j=1}^N f_j(x) g_j(y)$, $N\in \N$, where $f_j,g_j$ are 
bounded and continuous, is dense in the space of bounded continuous
functions w.r.t. the supremum norm. Since the measures $\nu_n$ tend to zero
weakly and $\sup_{n\in \N} \parallel \nu_n \parallel < \infty$,
this implies that the product measures $\nu_n \otimes \nu_n$ tend to zero 
weakly, too. Hence by (\ref{1.33}), we have proved (\ref{1.32a}) and 
therefore also (\ref{1.21}). 

It only remains to prove (\ref{1.22}). For this purpose we first note
that 
\begin{eqnarray}
c_n:= \int \int \vert k_{\nu_{nx}} (a) \vert^2 \, da \, \vert m \vert(dx) 
\longrightarrow 0 \qquad \mbox{as } n\longrightarrow \infty. \nonumber
\end{eqnarray}
This can be shown by mimicking the proof of (\ref{1.32a}). By (\ref{1.31}),
it follows that 
\begin{eqnarray}
\int  \vert    (G^{\nu_n} h \tilde{)}  (x) \vert^2 \, \vert m \vert (dx)
\le 2 c_n \parallel h \parallel_{H^2}^2,\qquad h\in H^2(\R^d). \nonumber
\end{eqnarray}
Thus, in order to prove (\ref{1.22}), we have only to show that there 
exists a finite constant $C$ such that 
\begin{eqnarray}\label{1.36}
\parallel G^m (I+G^m)^{-1} h \parallel_{L^2(dx)}
\le C  \left( \int \vert \tilde{h} \vert^2 d \vert m \vert \right)^{1/2},
\qquad h\in H^2(\R^d).  
\end{eqnarray}
Using the estimates (\ref{1.18d}), we have
\begin{eqnarray}\label{1.37}
G^m (I+G^m)^{-1} = - \sum_{j=1}^{\infty} (-G^m)^j.
\end{eqnarray}
According to (\ref{G5}),
$$ 
\parallel (G^m)^{j+1} h \parallel_{L^2(dx)} \le c 
\left( \int \vert \widetilde{ (G^m)^j h } \vert^2 d\vert m \vert \right)^{1/2} 
\le c\cdot c^j  
\left( \int \vert \tilde{h} \vert^2 d\vert m \vert \right)^{1/2},$$
for every $j\in \N$ and hence 
$$ 
\parallel \sum_{j=1}^{\infty} (-G^m)^j h \parallel_{L^2(dx)}
\le \sum_{j=1}^{\infty} c^j 
\left( \int \vert \tilde{h} \vert^2 d\vert m \vert \right)^{1/2}
= \frac{c}{1-c} \, \left(\int \vert \tilde{h} \vert^2 d\vert m \vert 
\right)^{1/2}.
$$
By (\ref{1.37}), this implies (\ref{1.36}) and the proof of the theorem
is complete. 
\hfill $\Box$

\begin{remark}
{\em We have shown that 
\begin{eqnarray*}
& \parallel (-\gD + \ee^2 \gD^2 + \mu_n + \alpha)^{-1} -
(-\gD + \ee^2 \gD^2 + m + \alpha)^{-1} \parallel^2 \\
\le & C_1 \int \int \vert \int g_{0,1}(y-a) g_{\ee,\alpha} (x-y) 
(m-\mu_n)(dy) \vert^2 da \, dx \\
& + C_2 
\int \int \vert \int g_{0,1}(y-a) g_{\ee,\alpha} (x-y) 
(m-\mu_n)(dy) \vert^2 da \, \vert m \vert (dx) 
\end{eqnarray*}
for some finite constants $C_j= C_j(\ee,\alpha)$, $j=1,2$, which can be
computed with the aid of the proof of theorem~\ref{thm1}. Thus the proof
provides explicit upper bounds for the error one makes when one replaces
the operator 
$-\gD + \ee^2 \gD^2 + m$ by $ -\gD + \ee^2 \gD^2 + \mu_n$. }
\end{remark} 

\begin{remark}\label{rem:spectrum}
{\em The essential spectrum of $-\gD + \ee^2 \gD^2 + m$ remains the same 
for any finite real-valued Radon measure $m$ on $\R^d$ 
$$ 
\sigma_{ess}(-\gD + \ee^2 \gD^2 + m) = 
\sigma_{ess}(-\gD + \ee^2 \gD^2 ) = [0,\infty). 
$$ 
By Sobolev's inequality and \cite[Lemma 19]{bpota},
the mapping $f\mapsto \tilde{f}$ from $H^2(\R^d)$ to $L^2(\vert m \vert)$ 
is compact. Therefore using estimate (\ref{G3}), one may conclude that
$G_{\ee,\alpha}^{\mu}$ is compact if regarded as an operator from
$H^2(\R^d)$ to $L^2(dx)$. According to the resolvent formula (\ref{1.17}), this
implies that the resolvent difference $G_{\ee,\alpha}^{m}-G_{\ee,\alpha}$ 
is compact and hence the corresponding essential spectra coincide.
}
\end{remark}


\section{Dependence on the coupling constant}\label{sec:lime}

In this section we are going to prove that 
\begin{eqnarray}\label{3.2}
-\gD + \ee^2 \gD^2 + m \longrightarrow -\gD + m
\qquad \mbox{as } \ee \downarrow 0,
\end{eqnarray}
in the norm resolvent sense. 
Here $m$ denotes a real-valued Radon measure on $\R^d$ and
we assume, in addition, 
that for every $\eta >0$ there exists a $\beta_{\eta} < \infty$ 
such that 
\begin{eqnarray}\label{3.1}
\int \vert f \vert^2 d\vert m \vert \le \eta \left( \int \vert \nabla f \vert^2 
dx + \beta_{\eta} \int \vert f \vert^2 dx \right), 
\qquad f\in C_0^{\infty}(\R^d).
\end{eqnarray}
Note that we neither require that $m$ is finite nor that $d\le 3$. 
On the other hand, the condition (\ref{3.1})
implies that $m(B)=0$ for every Borel set $B$ with classical capacity zero
and, for instance, it is excluded that $m$ is a point measure if $d>1$.

The inequality (\ref{3.1}) holds, in particular, provided $m$ belongs 
to the Kato class, i.e.
\begin{eqnarray*}
\sup_{n \in \Z} \vert m \vert ([n,n+1])&  < & \infty,\qquad d=1, \\
\lim_{\varepsilon \to 0} \sup_{x \in \R^2} \int_{B(x,\varepsilon)}
|\log(|x-y|)|\, \vert m \vert (dy) &=& 0, \qquad d=2, \\
\lim_{\varepsilon \to 0} \sup_{x \in \R^3} \int_{B(x,\varepsilon)}
\frac{1}{|x-y|}\, \vert m \vert (dy) &=& 0, \qquad d=3, 
\end{eqnarray*}
with $B(x,\varepsilon)$ denoting the ball of radius $\varepsilon$ 
centered at $x$ (cf. \cite{sv}, Theorem 3.1). We refer to 
\cite[chapter~1.2]{cfks}, for additional examples of measures
satisfying (\ref{3.1}). 

In general, the elements $f$ in the form domain of $-\gD$ do not possess
a continuous representative $\tilde{f}$. Therefore we shall give a 
definition of $\kE_{\ee}^m$ different from the one in section \ref{sec:mu}
so that it works for all $\ee \ge 0$. Of course, both definitions are 
equivalent in the special case of positive $\ee$. 

Since the space $C_0^{\infty}(\R^d)$ of smooth functions with compact support 
is dense in the Sobolev space $H^1(\R^d)$, there exists a unique
bounded linear mapping $J_m:H^1(\R^d) \longrightarrow L^2(\vert m \vert)$ 
satisfying 
\begin{eqnarray}
J_m f =f, \qquad f\in  C_0^{\infty}(\R^d), \nonumber
\end{eqnarray}
(strictly speaking $J_m$ maps the $dx$-equivalence class of the
continuous function $\tilde{f}\in  C_0^{\infty}(\R^d)$ 
to the $\vert m \vert$-equivalence class of $\tilde{f}$). 
We put
\begin{eqnarray}
D(\kE_{\ee}^m) & := & D(\kE_{\ee}),\nonumber \\
\kE_{\ee}^m (f,f) & := &  \kE_{\ee} (f,f) + 
(A_m J_m f ,J_m f)_{L^2(\vert m \vert)},
\qquad f\in D(\kE_{\ee}^m), \nonumber
\end{eqnarray}
where $D(\kE_{\ee})=H^1(\R^d)$ for $\ee = 0$, $D(\kE_{\ee})=H^2(\R^d)$
otherwise and
\begin{eqnarray*}
A_m h (x) := \left \{ 
\begin{array}{ll}
h(x),& \quad x\in B,\\
-h(x), & \quad x\in \R^d\setminus B,
\end{array} \right. \qquad 
h\in L^2(\vert m \vert),
\end{eqnarray*}
with $B$ being any Borel set such that $m^+(\R^d\setminus B) = 0 = m^- (B)$. 
By (\ref{3.1}) and the KLMN-theorem,  
the quadratic form $\kE_{\ee}^m$ in $L^2(dx)$ is lower 
semibounded and closed and 
\begin{eqnarray}
\kE_{\ee,\beta_{1}}^m (f,f) \ge 0, \qquad f\in D(\kE_{\ee}^m). \nonumber
\end{eqnarray}
Again, $-\gD + \ee^2 \gD^2 +m$ denotes the lower semibounded self-adjoint
operator associated to $\kE_{\ee}^m$ and we put
$$
R_{\ee,\alpha}^m:= (-\gD + \ee^2 \gD^2 + m + \alpha)^{-1}
$$
provided the inverse operator exists. $G_{\ee,\alpha}$ is defined the same way
as in section \ref{sec:mu}.

One key for the proof of the convergence result (\ref{3.2}) 
is the observation that one can decompose 
\begin{eqnarray}
\hat{g}_{\ee,\alpha} (p) = 
\frac{c(\ee)}{p^2 + \alpha(\ee)} - \frac{c(\ee)}{p^2 + \beta(\ee)}, \nonumber
\end{eqnarray}
whenever $c(\ee)$ is defined. The coefficients $-\alpha(\ee)$ and 
$-\beta(\ee)$ are the roots of the polynomial $\ee^2 x^2 + x + \alpha$;
a simple calculation yields
\begin{eqnarray}
c(\ee) &:=& \frac{1}{\sqrt{1- 4 \ee^2 \alpha}} \qquad \longrightarrow 1,
 \qquad \mbox{as } \ee \downarrow 0, \nonumber \\
\alpha(\ee) &:=& \frac{2\alpha}{1+ \sqrt{1-4\ee^2 \alpha}} 
\longrightarrow \alpha,
\qquad \mbox{as } \ee \downarrow 0,
\label{3.4}\\
\beta(\ee) &:=&  \frac{1+ \sqrt{1-4\ee^2 \alpha}}{2\ee^2}
\longrightarrow \infty,
\qquad \mbox{as } \ee \downarrow 0 \nonumber.
\end{eqnarray}
Using the parameters introduced above, we arrive at
\begin{eqnarray}\label{3.9}
G_{\ee,\alpha} = c(\ee) G_{0,\alpha(\ee)} - c(\ee) G_{0,\beta(\ee)}.
\end{eqnarray}
%


In the proof of the convergence result (\ref{3.2}) we will use again 
a Krein-like resolvent formula, this time using the one from \cite{bpeking}, 
cf.(\ref{3.23}) below. First we need some preparation.
Let $\alpha >0$ and $\ee \ge 0$. We introduce the operator 
$J_{m,\ee,\alpha}$ from the Hilbert space $(D(\kE_{\ee}), \kE_{\ee,\alpha})$ 
to $L^2(\vert m \vert )$ as follows: 
\begin{eqnarray}
D(J_{m,\ee,\alpha}) & := & D(\kE_{\ee}),\nonumber \\
J_{m,\ee,\alpha} f & := & J_m f, \qquad f\in D(J_{m,\ee,\alpha}). \nonumber
\end{eqnarray}
By (\ref{3.1}), the operator norm of $J_{m,\ee,\alpha}$ is less
than or equal to $\eta$ provided $\alpha \ge \beta_{\eta}$. 
Thus we can choose $\alpha_0> 0$ and $c<1$ such that 
\begin{eqnarray}\label{3.22}
\parallel J_{m,\ee,\alpha} 
\parallel_{(D(\kE_{\ee}), \kE_{\ee,\alpha}), L^2(\vert m \vert)}
\, \le \sqrt{c},\qquad \alpha\ge \alpha_0.
\end{eqnarray}

Due to (\ref{3.22}), the hypothesis of Theorem 3 in \cite{bpeking} is 
satisfied 
and the theorem implies that $-\alpha$ belongs to the resolvent set
of $ -\gD + \ee^2 \gD^2 +m$ and
\begin{eqnarray}\label{3.23}
R_{\ee,\alpha}^m = G_{\ee,\alpha} - 
(J_{m,\ee,\alpha})^* A_m  (1+ J_m J_{m,\ee,\alpha}^* A_m)^{-1} 
J_m G_{\ee,\alpha},\qquad \alpha \ge \alpha_0.
\end{eqnarray}
In fact, we can write
\begin{eqnarray}\label{3.43}
J_{m,\ee,\alpha'}^* = (J_m G_{\ee,\alpha'})^* ,\qquad \alpha' >0,
\end{eqnarray}
since we have 
$$ (J_{m,\ee,\alpha'}^* f, h) = 
\kE_{\ee,\alpha'} ( J_{m,\ee,\alpha'}^* f, G_{\ee,\alpha'} h) =
(f, J_{m,\ee,\alpha'}  G_{\ee,\alpha'} h)_{L^2(\vert m \vert)} =
( (J_m  G_{\ee,\alpha'})^* f , h)$$
for every  $ h\in L^2(dx)$, $\ee\ge 0$ and $\alpha' >0$. 

\begin{theorem}\label{th3.1}
Let $m$ be a real-valued Radon measure on $\R^d$ satisfying 
(\ref{3.1}). Then the operators $-\gD + \ee^2 \gD^2 +m$ converge
to $-\gD + m$ in the norm resolvent sense as $\ee\downarrow 0$. 
\end{theorem}
{\bf Proof}:
Both resolvents are written by means of Krein's formula (\ref{3.23}), 
so we can compare the first and second terms separately.  To see that
$\parallel G_{\ee,\alpha} - G_{0,\alpha} \parallel_{L^2(dx)}$ vanishes
in the limit $\ee \downarrow 0$ is simple. It is enough to use
the first resolvent formula,
\begin{eqnarray}\label{3.42}
G_{0,\alpha(\ee)} - G_{0,\alpha} = (\alpha - \alpha(\ee) ) G_{0,\alpha}
G_{0,\alpha(\ee)}
\end{eqnarray}
and the fact that 
\begin{eqnarray}
\parallel G_{0,\alpha'} \parallel_{L^2(dx),H^1}^2 \le k(\alpha'),\qquad
\alpha' >0, \nonumber
\end{eqnarray}
for some continuous function $k$ vanishing at infinity (actually,
$k(x) = 1/x^2$ for $x\le 2$ and $k(x) = 1/(4(x-1))$ for $x>2$).
Then the decomposition (\ref{3.9}) of $G_{\ee,\alpha}$ and the
asymptotic behavior 
(\ref{3.4}) of $\alpha(\ee), \beta(\ee)$ and $c(\ee)$ finish the argument.

The proof that also the difference of second terms in Krein's formula 
tend to zero as $\ee \to 0$ can be reduced into two tasks
\begin{eqnarray}
\parallel J_m G_{\ee,\alpha} - 
J_m G_{0,\alpha} \parallel_{L^2(dx),L^2(\vert m \vert)} &\longrightarrow& 0
\qquad \mbox{as } \ee \downarrow 0, \nonumber \\
\parallel  (1+ J_m J_{m,\ee,\alpha}^* A_m)^{-1} 
-   (1+ J_m J_{m,0,\alpha}^* A_m)^{-1} \parallel_{L^2(\vert m \vert)} 
&\longrightarrow& 0 \qquad \mbox{as } \ee \downarrow 0. \nonumber
\end{eqnarray}
The argument for the first line is similar to the one we have presented
above for $G_{\ee,\alpha} - G_{0,\alpha}$, we only have to add that, by
hypothesis (\ref{3.1}), it follows that
\begin{eqnarray}\label{3.41}
\parallel J_m G_{0,\alpha'} \parallel_{L^2(dx),L^2(\vert m \vert)}^2 
\le \mbox{max}(1,\beta_1) k(\alpha'),\qquad
\alpha' >0, 
\end{eqnarray}
where function $k(\alpha')$ is defined as above.

To show the second line we choose any $\alpha > \alpha_0$, then from 
(\ref{3.22}) we get 
\begin{eqnarray}
\parallel  (1+ J_m J_{m,\ee,\alpha}^* A_m)^{-1}\parallel_{L^2(\vert m \vert)}
\le \frac{1}{1-c},\qquad \ee\ge 0. \nonumber 
\end{eqnarray}
By the second resolvent identity
$$ 
(1+A)^{-1} - (1+B)^{-1} = (1+A)^{-1} (B-A) (1+B)^{-1},
$$
it is sufficient to prove that
\begin{eqnarray}\label{3.28}
\parallel  J_m J_{m,\ee,\alpha}^*  -  J_m J_{m,0,\alpha}^*
\parallel_{L^2(\vert m \vert)}
\longrightarrow 0 \qquad \mbox{as } \ee \downarrow 0.
\end{eqnarray}
From (\ref{3.9}) and (\ref{3.43}) follows that
\begin{eqnarray}
J_m J_{m,\ee,\alpha}^* = 
c(\ee) J_m     (J_m  G_{0,\alpha(\ee)})^*  -
c(\ee) J_m  (J_m  G_{0,\beta(\ee)})^* , \nonumber
\end{eqnarray}
note that $c(\ee)$ is real for sufficiently small $\ee$.
Using this expression and (\ref{3.42}) and (\ref{3.43}), we get
\begin{eqnarray}
& \parallel J_m J_{m,\ee,\alpha}^* - J_m J_{m,0,\alpha}^* 
\parallel_{L^2(\vert m \vert)} \nonumber \\
\le &  \parallel ( c(\ee)-1) 
J_m (J_m G_{0,\alpha(\ee)})^* 
\parallel_{L^2(\vert m \vert)} 
+ \parallel 
J_m (J_m G_{0,\alpha(\ee)})^* - J_m (J_m G_{0,\alpha})^*
\parallel_{L^2(\vert m \vert)} 
\nonumber \\
& + \parallel c(\ee)
J_m (J_m G_{0,\beta(\ee)})^*
\parallel_{L^2(\vert m \vert)} \nonumber \\
=  &  \parallel ( c(\ee)-1) 
J_{m,0,\alpha(\ee)} J_{m,0,\alpha(\ee)}^* 
\parallel_{L^2(\vert m \vert)} 
+ \parallel  (\alpha - \alpha(\ee) ) 
J_m G_{0,\alpha} (J_m G_{0,\alpha(\ee)})^* \parallel_{L^2(\vert m \vert)} 
\nonumber \\
& + \parallel c(\ee)
J_{m,0,\beta(\ee)} J_{m,0,\beta(\ee)}^*
\parallel_{L^2(\vert m \vert)},\qquad \ee >0.  \nonumber
\end{eqnarray}
According to (\ref{3.1}), the mapping 
$\parallel J_{m,0,\alpha} J_{m,0,\alpha}^* \parallel_{L^2(\vert m \vert)}$
is locally bounded for $\alpha \in (0,\infty)$ and tends to zero as $\alpha$ 
tends to infinity. Since $\alpha(\ee) \longrightarrow \alpha$,
$c(\ee) \longrightarrow 1$ 
and $\beta(\ee) \longrightarrow \infty$ as $\ee \downarrow 0$, 
this implies, in conjunction with (\ref{3.41}), that (\ref{3.28}) holds. 
\hfill $\Box$

\begin{remark}\label{th3.2}
{\em By the proof above, 
$\parallel G_{\ee,\alpha}^m - G_{0,\alpha}^m \parallel$ is upper bounded
by an expression of the form $c\cdot ( \ee^2 + \eta(m,\ee) )$ where
the finite constant $c$ can be extracted from the proof and
$\eta(m,\ee)$ has to be chosen (and can be chosen) such that (\ref{3.1}) 
holds with $\eta$ and $\beta$ replaced by $\eta(m,\ee)$ and $\beta(\ee)$, 
respectively.} 
\end{remark}


\section{Eigenvalues and eigenspaces of the approximating operators}
\label{sec:eig}

Throughout this section let $d\le 3$ and let $m$ be a finite real-valued 
Radon measure satisfying (\ref{3.1})  (e.g., let $m$ be from the Kato class).
By the two preceding convergence results, we can approximate the operator
$-\gD +m$ in $L^2(\R^d,dx)$ by operators of the form $-\gD + \ee^2 \gD^2 +\mu$,
where $\ee>0 $ and $\mu$ is a point measure with mass at only finitely 
many points. Since the convergence is in norm resolvent sense, we can 
thus approximate the negative eigenvalues and corresponding eigenspaces of 
the former operator by the corresponding eigenvalues and eigenfunctions 
of the latter one. Note that we know from remark \ref{rem:spectrum} and
\cite[Theorem 3.1]{beks} that the essential spectra coincide.

The following theorem shows how to compute the eigenvalues 
and corresponding eigenspaces of the approximating operators.

\begin{theorem}\label{th3.2a}
Let $d\le 3$ and $\ee >0$. 
Let $\mu = \sum_{j=1}^N c_j \delta_{x_j}$, where $N\in \N$, 
$x_1,\ldots, x_N$ are $N$ distinct points in $\R^d$ and 
$c_1,\ldots,c_N$ are real numbers different from zero.
Then the following holds:

\noindent
a) The real number $-\alpha < 0$ is an eigenvalue of 
$-\gD + \ee^2 \gD^2 + \mu$ if and only if 
$$
\det \left( \frac{\delta_{jk}}{c_k} +  g_{\ee,\alpha}(x_j - x_k) 
\right)_{1\le j,k \le N} = 0.
$$

\noindent
b) For every eigenvalue $-\alpha <0$ the corresponding eigenfunctions have
the following form 
$$  
\sum_{k=1}^N h_k g_{\ee,\alpha} (\cdot - x_k), \qquad
(h_k)_{1\le k \le N}^T \in \ker 
\left(\frac{\delta_{jk}}{c_k} + g_{\ee,\alpha}(x_j - x_k) 
\right)_{1\le j,k \le N}
$$
\end{theorem}
{\bf Proof}:
Since $D(\kE_{\ee})=H^2(\R^d)$, the mapping $J_{\mu}$
can be understood as
$$ 
J_{\mu} f := \tilde{f} \qquad \vert \mu \vert 
\mbox{-a.e.},\quad f\in H^2(\R^d).
$$
By (\ref{1.10a}), $\int g_{\ee,\alpha} (\cdot -y) f(y)dy $
is the unique continuous representative of $G_{\ee,\alpha}f$. Hence
$J_{\mu} G_{\ee,\alpha}$ is the integral operator from $L^2(dx)$
to $L^2(\vert \mu \vert)$  with kernel $g_{\ee,\alpha}(x-y)$ and
its inverse operator $(J_{\mu} G_{\ee,\alpha})^*$
is the integral operator from $L^2(\vert \mu \vert)$ to $L^2(dx)$
with the same kernel. Thus we get
\begin{eqnarray}\label{3.61}
J_{\mu}  (J_{\mu} G_{\ee,\alpha})^*  A_{\mu} h(x_j) = 
 \sum_{k=1}^N c_k g_{\ee,\alpha} (x_j - x_k) h(x_k),\qquad 1\le j \le N,
\end{eqnarray}
for every $h\in L^2(\vert \mu \vert)$.

Due to Krein's formula (\ref{3.23}), $-\alpha<0$ belongs to the resolvent
set of $(-\gD + \ee^2 \gD^2 + \mu)$ provided $1+ J_{\mu}(J_{\mu} G_{\ee,\alpha}
)^* A_{\mu}$ is bijective. Since $L^2(\vert \mu \vert )$ is finite 
dimensional and we have expression (\ref{3.61}), that is true if and only if 
\begin{eqnarray}
\lambda(\alpha) := \mbox{det} (\delta_{jk} + 
c_k g_{\ee,\alpha}(x_j - x_k) )_{1\le j,k \le N} \not = 0, \nonumber
\end{eqnarray}
with $\delta_{j,k}$ being the Kronecker delta. 
As $g_{\ee,\alpha} (x)$ is a real analytic function of $\alpha \in (0,\infty)$
for every $x\in \R^d$, the function $\lambda(\alpha)$
is also real analytic on $(0,\infty)$. By (\ref{G2}), it is different from
zero for all sufficiently large $\alpha$. Thus the set of zeros on
$(0,\infty)$ of this function is discrete. 

Since $J_{\mu}G_{\ee,\alpha}$ is surjective and 
$(J_{\mu}G_{\ee,\alpha})^* A_{\mu} $ injective, the resolvent formula
(\ref{3.23}) implies that any $\alpha_0 >0$ satisfying $\lambda(\alpha_0) = 0$
is a pole of $R^{\mu}_{\ee,\alpha}$.
Thus we have proved that $-\alpha_0$ is an eigenvalue of 
$-\gD + \ee^2 \gD^2 + \mu$ if and only if $\lambda(\alpha_0)=0$. Finally,
the expression
$$
\det(\delta_{jk} + c_k g_{\ee,\alpha}(x_j - x_k) )_{1\le j,k \le N}
= \Pi_{k=1}^N c_k \cdot
 \det(\delta_{jk}/c_k +  g_{\ee,\alpha}(x_j - x_k) )_{1\le j,k \le N}
$$
implies the assertion a).

By the preceding considerations and \cite[Lemma 1]{bpeking}, 
$$ h\mapsto (J_{\mu} G_{\ee,\alpha})^* A_{\mu} h$$
is a linear bijective mapping from 
$\ker (1+ J_{\mu} (J_{\mu} G_{\ee,\alpha})^* A_{\mu})$ onto 
$\ker (-\gD + \ee^2 \gD^2 +\mu + \alpha)$. The assertion b) follows
from a simple algebraic calculation. 
\hfill $\Box$

\begin{remark}\label{th3.3}
{\em 
Since the Hilbert space $L^2(\vert \mu \vert)$ is $N$-dimensional with 
$N< \infty$, the resolvent formula (\ref{3.23}) implies that the difference 
$G_{\ee,\alpha}^{\mu} - G_{\ee,\alpha}$ is a finite rank operator with
rank less than or equal to $N$. Thus 
the number, counting multiplicity,
 of negative eigenvalues
of $-\gD + \ee^2 \gD^2 + \mu$ is less than or equal to $N$. }
\end{remark}


Let us illustrate the approximation by point measures on a simple
example in dimension two. Suppose that measure $m$ is minus length measure 
supported by a circle of radius $R$, i.e. $m$ is constant and negative measure. 
This makes the choice of approximating point measures very simple: we spread 
equidistantly $N$ points along the circle and all the points have the same 
coupling constant $c$
$$
c= - \frac{\gamma 2 \pi R} {N}.
$$

\begin{figure}[!b]
\centering
\includegraphics[angle=-90, width=\textwidth]{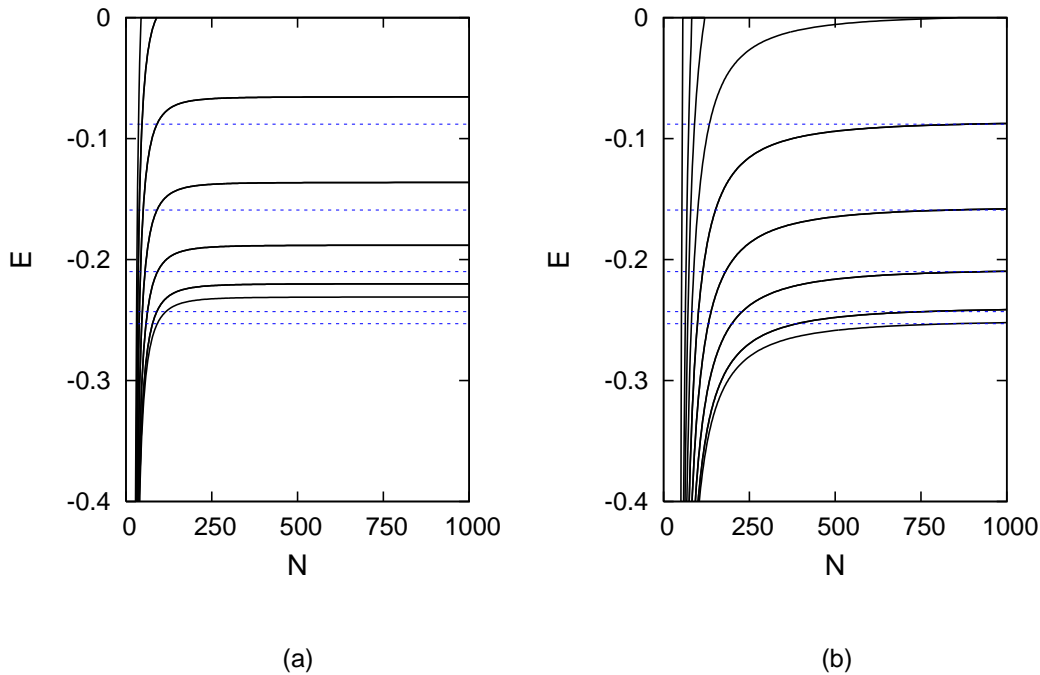}
\vspace{-1cm}
%
%
\caption{The dependence of the approximate eigenvalues on the number 
of point potentials for circle with $R=10$ and $\ee=0.1$ (a), $\ee=0.01$ (b).
The dashed lines represent the exact eigenvalues of $-\gD +m$.}
\label{fig:spec1}
\end{figure}

Due to the symmetry, the spectrum of $-\Delta + m$ for this specific measure
is known; it consists of the essential 
spectrum $[0,\infty)$ and a finite number of negative eigenvalues, which are
all except the lowest one twice degenerate, see \cite{ags}. 
To find the eigenvalues, one has to decompose $L^2(\R^2)$ into
angular momentum subspaces and then to look for solutions of an implicit 
equation in each of the subspaces. Therefore we can compute and compare both 
exact and approximate eigenvalues.

Each approximation is characterized by a pair of numbers, $\ee>0$ and $N
\in \N$. In numerical calculations we fix $\ee$ and we let $N$ grow. The
results for one chosen radius and two different parameters $\ee$ are
depicted in figure~\ref{fig:spec1}, cases (a) and (b) correspond to
$\ee=0.1$ and $\ee=0.01$, respectively.  
We observe that below some threshold number of points, the approximate discrete
spectrum has no resemblance to the exact spectrum. The approximate eigenvalues 
may be very large negative and their number may be much higher than the
number of exact eigenvalues (in figure~\ref{fig:spec1}, we even have not plotted
all the eigenvalues which exist only for small $N$.)

It appears that for larger $\ee$, we get a fast convergence of eigenvalues, 
however, they are all shifted from the exact ones. The reason is that since we 
work with fixed $\ee$, the limit operator is in fact $-\gD +\ee^2\gD^2 +m$ instead
of $-\gD + m$. On the contrary, small $\ee$ means that one need more points to
obtain a qualitatively correct spectrum, but then for a large number of points
one gets much closer to the exact spectrum.

\begin{figure}[!t]
\centering
\includegraphics[angle=-90, width=0.51\textwidth]{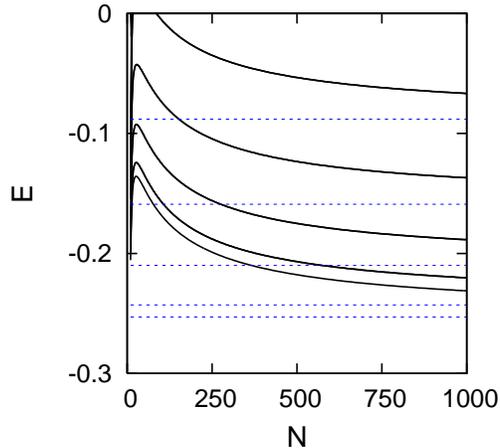}
%
%
\caption{The dependence of the approximate eigenvalues on the number 
of point potentials for $R=10$, using the standard two-dimensional point
potentials. The dashed lines represent the exact eigenvalues of $-\gD +m$.}
\label{fig:spec2}
\end{figure}

We can also compare this approximation to \cite{en}, where approximating
operators were Laplacians with point potentials. Those point potentials are
of course different, they are not defined via a quadratic form and cannot be
understood as a special case $\ee=0$ of section~\ref{sec:mu}, 
instead boundary conditons on wavefunctions are used, see 
\cite{aghh}. Figure~\ref{fig:spec2} presents the eigenvalues of Laplacians 
perturbed by point potentials which converge to $-\gD+m$ with the same 
measure $m$ as above. We have already mentioned in the introduction that here, 
we obtain a stronger convergence result than the one in \cite{en}. Moreover, 
comparing both figures \ref{fig:spec1} and \ref{fig:spec2}, we see that 
employing fourth-order differential operators in the approximation may improve 
significantly the spectral convergence.

\subsection*{Appendix}

In section \ref{sec:mu} we have employed Krein's formula (\ref{1.17}).
Various forms of this formula can be found in the literature. Let us prove
here the one we have used.
 
Let $f\in L^2(dx)$. Since $\kE_{\ee}$ and $\kE_{\ee}^{\mu}$
are associated to $-\gD + \ee^2 \gD^2$ and $-\gD + \ee^2\gD^2 + \mu$,
respectively, it follows from Kato's representation theorem that
\begin{eqnarray}\label{1.17c}
\kE_{\ee,\alpha} (G_{\ee,\alpha}f, h) = (f,h) = 
\kE_{\ee,\alpha}^{\mu} ((-\gD + \ee^2 \gD^2 + \mu + \alpha)^{-1} f, h),
\end{eqnarray}
for any $h \in H^2(\R^d)$ and $f \in L^2(\R^d)$. Moreover we have 
\begin{eqnarray}\label{1.17d}
& \kE_{\ee,\alpha} (G_{\ee,\alpha}^{\mu} \psi, h) = 
( G_{\ee,\alpha}^{\mu} \psi, (-\gD +\ee^2 \gD^2 + \alpha) h) \nonumber \\
= & \int \int g_{\ee,\alpha} (x-y) \bar{ \tilde{\psi} } (y) \mu(dy) 
(-\gD +\ee^2 \gD^2 + \alpha) h (x) \,dx \nonumber \\
= & \int \int g_{\ee,\alpha} (x-y) (-\gD +\ee^2 \gD^2 + \alpha) h (x) \,dx
\, \bar{ \tilde{\psi} } (y) \mu(dy) \nonumber \\
= & \int \tilde{h} \bar{ \tilde{\psi} } \,\mu(dy), \qquad 
\psi \in H^2(\R^d),\quad 
h\in D(-\gD + \ee^2 \gD^2). 
\end{eqnarray}
We could change the order of integration in the second step. 
In fact, as $\mu^{\pm}$ are finite Radon measures and $g_{\ee,\alpha}$ 
is square integrable w.r.t. the Lebesgue measure $dx$, 
the mappings 
$x\mapsto \int \vert g_{\ee,\alpha} (x-y) \vert \mu^{\pm}(dy)$, 
$\R^d \longrightarrow \R$, are square integrable w.r.t.
$dx$. Since $\tilde{\psi}$ is bounded and 
$(-\gD + \ee^2 \gD^2 + \alpha)h\in L^2(dx)$ it follows that 
$$  \int \int \vert g_{\ee,\alpha} (x-y) \bar{ \tilde{\psi} } (y) \vert
\mu^{\pm}(dy) 
\vert  (-\gD +\ee^2 \gD^2 + \alpha) h (x) \vert dx <\infty $$
and, by Fubini's theorem, we could change the order of integration in the
second step. In the last step we have used (\ref{1.10a}). Employing Sobolev's
inequality and the fact that
$D( -\gD + \ee^2 \gD^2)$ is dense in $( D( \kE_{\ee}), \kE_{\ee,\alpha})$,
we can extend (\ref{1.17d}) to all functions $\psi, \, h\in D(\kE_{\ee})$.

Put 
\begin{eqnarray*}
\phi:= G_{\ee,\alpha} f - G_{\ee,\alpha}^{\mu} 
( I+ G_{\ee,\alpha}^{\mu})^{-1} G_{\ee,\alpha} f.
\end{eqnarray*}
Then $\phi\in H^2(\R^d)= D(\kE_{\ee}^{\mu})$ and (\ref{1.17c}) and 
extended (\ref{1.17d}) yield
\begin{eqnarray}
\kE_{\ee,\alpha}^{\mu} (\phi, h)  &=& \kE_{\ee,\alpha} (G_{\ee,\alpha} f, h) 
- \kE_{\ee,\alpha} (G_{\ee,\alpha}^{\mu} (I+ G_{\ee,\alpha}^{\mu})^{-1}
G_{\ee,\alpha} f,h) \nonumber \\
& & +   \int [ G_{\ee,\alpha}f - G_{\ee,\alpha}^{\mu} 
(I+ G_{\ee,\alpha}^{\mu})^{-1} 
G_{\ee,\alpha} f \bar{\tilde{]}} \tilde{h} d\mu   \nonumber \\
&=& (f,h) - 
\int  [ (I+ G_{\ee,\alpha}^{\mu})^{-1} G_{\ee,\alpha} f \bar{\tilde{]}}  
\tilde{h} d\mu \nonumber \\
& & + \int [ (I+ G_{\ee,\alpha}^{\mu}) (I+ G_{\ee,\alpha}^{\mu})^{-1}
G_{\ee,\alpha}f - G_{\ee,\alpha}^{\mu} (I+ G_{\ee,\alpha}^{\mu})^{-1} 
G_{\ee,\alpha} f \bar{\tilde{]}} \tilde{h} d\mu\nonumber \\
&=&  (f,h),\qquad h\in H^2(\R^d). \nonumber
\end{eqnarray}
Due to (\ref{1.17a}), $\kE_{\ee,\alpha}^{\mu}$ is a scalar product
on $D(\kE_{\ee,\alpha}^{\mu})=  H^2(\R^d)$. 
Thus (\ref{1.17c}) and the calculation above imply that
$\phi = (-\gD + \ee^2 \gD^2 + \mu + \alpha)^{-1} f$.

\subsection*{Acknowledgment}
 
This work is partially supported by the Marie Curie fellowship 
MEIF-CT-2004-009256.

\end{document}